# Secure multi-party quantum summation based on quantum Fourier transform

Hui-Yi Yang, Tian-Yu Ye*

College of Information & Electronic Engineering, Zhejiang Gongshang University, Hangzhou 310018, P.R.China
*E-mail：happyyty@aliyun.com

**Abstract:** In this paper, we propose a novel secure multi-party quantum summation protocol based on quantum Fourier transform, where the traveling particles are transmitted in a tree-type mode. The party who prepares the initial quantum states is assumed to be semi-honest, which means that she may misbehave on her own but will not conspire with anyone. The proposed protocol can resist both the outside attacks and the participant attacks. Especially, one party cannot obtain other parties' private integer strings; and it is secure for the colluding attack performed by at most $n-2$ parties, where $n$ is the number of parties. In addition, the proposed protocol calculates the addition of modulo $d$ and implements the calculation of addition in a secret-by-secret way rather than a bit-by-bit way.

## 1 Introduction

Quantum cryptography, which can be regarded as the combination of quantum mechanics and classical cryptography, has attracted a lot of attention since it was derived by Bennett and Brassard [1] in 1984, as it can attain unconditional security in theory through the physical principles of quantum mechanics. During the past three decades, quantum cryptography was widely investigated so that numerous branches have been established, such as quantum key distribution (QKD) [1-5], quantum secure direct communication (QSDC) [6-8], quantum secret sharing (QSS) [9-11], quantum key agreement (QKA) [12-40], quantum private query (QPQ) [41-45] *etc*.

Secure multi-party computation, first introduced by Yao [46] and extended by Goldreich *et al.* [47], is a significant subfield of classical cryptography. Naturally, whether the physical principle of quantum mechanics can be applied into secure multi-party computation is an important and interesting question. To date, many researchers have investigated secure multi-party computation within quantum settings [48-51]. Lo [48] thought that the equality function cannot be securely evaluated in a two-party scenario. Thus, some additional assumptions, such as a third party (TP), should be considered. Ben-Or *et al.* [49] studied the question that in order for distributed quantum computations to be possible, how many players must keep honest. Chau [50] put forward a scheme to improve the speed of classical multi-party computation with quantum techniques. Smith [51] pointed out that any multi-party quantum computation can be secure as long as the number of dishonest players is less than $n/6$.

Secure multi-party summation, which can be used to construct complex secure protocols for other multi-party computation, is a fundamental problem of secure multi-party computation. It can be formulated as follows [52]: $n$ players, $P_1, P_2, \ldots, P_n$, want to evaluate a summation function $f(x_1, x_2, \ldots, x_n)$, where $x_i$ is the secret value from $P_i$. The result of this function can be revealed publicly or privately to some particular player. The task of secure multi-party summation is to preserve the privacy of the players' inputs and guarantee the correctness of computation. In 2002, Heinrich [53] investigated quantum summation with an application to integration. In 2003, Heinrich [54] studied quantum Boolean summation with repetitions in the worst-average setting. In 2006, Hillery [55] put forward a multi-party quantum summation protocol by using two-particle $N$-level entangled states which accomplishes the summation of $N$ players in voting procedure on the basis of ensuring the anonymity of players. In 2007, Du *et al.* [56] suggested a novel scheme of secure quantum addition modulo $n+1$ $(n \geq 2)$ by using non-orthogonal states, which can add a number to an unknown number secretly. Here, $n$ represents the number of parties carrying a secret. In 2010, Chen *et al.* [52] proposed a quantum addition modulo 2 protocol based on multi-particle GHZ entangled states. In 2014, Zhang *et al.* [57] constructed a high-capacity quantum addition modulo 2 protocol with single photons in both polarization and spatial-mode degrees of freedom. In 2015, Zhang *et al.* [58] suggested a three-party quantum addition modulo 2 protocol by using six-qubit genuinely maximally entangled states. In 2016, Shi *et al.* [59] thought that the protocols in Refs.[52,56] have two drawbacks: one the one hand, the modulo of these two protocols is too small, resulting in the limitation for more extensive applications; on the other hand, these two protocols do not possess an enough high computation efficiency because of their bit-by-bit computation. Then, they proposed a quantum addition modulo $N$ protocol through quantum Fourier transform, controlled-not operation, oracle operation and inverse quantum Fourier transform, which implements the calculation of summation in a secret-by-secret way rather than a bit-by-bit way. Here, $N=2^m$ and $m$ is the number of qubits represented by one basis state. In this protocol, the calculations of secure multi-party summation are securely transferred into the calculations of the corresponding phase information by quantum Fourier transform. And later, the phase information is extracted after an inverse quantum Fourier transform. In 2017, Shi and Zhang [60] presented a common quantum solution to a class of special two-party private summation problems. In the same year, Zhang *et al.* [61] put forward a multi-party quantum addition modulo 2 protocol without a trusted TP based on single particles.

Based on the above analysis, in this paper, we propose a novel secure multi-party quantum summation protocol based on quantum Fourier transform. The party who prepares the initial quantum states is assumed to be semi-honest, which means that she may misbehave on her own but will not conspire with anyone. The proposed protocol can resist both the outside attacks and the participant attacks. Especially, one party cannot obtain other parties' private integer strings; and it is secure for the colluding attack performed by at most $n-2$ parties. In addition, the proposed protocol calculates the addition of modulo $d$, and implements the calculation of addition in a secret-by-secret way rather than a bit-by-bit way.

The rest of this paper is organized as follows. In Sect.2, we introduce the preliminary knowledge used in this paper. In Sect.3, we describe and analyze the proposed secure multi-party quantum summation protocol. Finally, discussion and conclusion are given in Sect.4.

## 2 Preliminary knowledge

Before depicting the proposed protocol, it is necessary for us to introduce the preliminary knowledge first.

### 2.1 Quantum Fourier transform and its application

Let us define the $d$-level $n$-particle entangled state as follows:

$$|\omega\rangle_{12...n} = \frac{1}{\sqrt{d}} \sum_{r=0}^{d-1} |r\rangle_1 |r\rangle_2 ... |r\rangle_n, \quad (1)$$

where each $|r\rangle$ is a $d$-level basis state, $r \in \{0,1,...,d-1\}$ and $d \geq 2$. For each $d$-level basis state $|r\rangle$, the $d$th order discrete quantum Fourier transform is defined to be

$$F|r\rangle = \frac{1}{\sqrt{d}} \sum_{l=0}^{d-1} \zeta^{lr} |l\rangle, \quad (2)$$

where $\zeta = e^{2\pi i/d}$. The two sets, $V_1 = \{|r\rangle\}_{r=0}^{d-1}$ and $V_2 = \{F|r\rangle\}_{r=0}^{d-1}$, are two common conjugate bases.

Further, we define a transformation operation $U_k$ as follows:

$$U_k = \sum_{u=0}^{d-1} |u \oplus k\rangle\langle u|, \quad (3)$$

where $k$ runs from 0 to $d-1$. Throughout this paper, $\oplus$ represents the addition modulo $d$. Apparently, after the operation $U_k$ is performed on the $d$-level basis state $|r\rangle$, we can obtain

$$U_k |r\rangle = |r \oplus k\rangle. \quad (4)$$

After performing the operation $(U_{k_1}F) \otimes (U_{k_2}F) \otimes ... \otimes (U_{k_n}F)$ ( $k_1, k_2, ..., k_n \in \{0,1,...,d-1\}$ ) on the state $|\omega\rangle_{12...n}$, we can get

$$\begin{aligned}
&(U_{k_1}F) \otimes (U_{k_2}F) \otimes ... \otimes (U_{k_n}F) |\omega\rangle_{12...n} \\
&= \frac{1}{\sqrt{d}} \sum_{r=0}^{d-1} (U_{k_1}F)|r\rangle_1 \otimes (U_{k_2}F)|r\rangle_2 \otimes ... \otimes (U_{k_n}F)|r\rangle_n \\
&= \frac{1}{\sqrt{d}} \sum_{r=0}^{d-1} \left( U_{k_1} \frac{1}{\sqrt{d}} \sum_{l_1=0}^{d-1} \zeta^{l_1 r} |l_1\rangle \right) \otimes \left( U_{k_2} \frac{1}{\sqrt{d}} \sum_{l_2=0}^{d-1} \zeta^{l_2 r} |l_2\rangle \right) \otimes ... \otimes \left( U_{k_n} \frac{1}{\sqrt{d}} \sum_{l_n=0}^{d-1} \zeta^{l_n r} |l_n\rangle \right) \\
&= d^{-\frac{n+1}{2}} \sum_{l_1, l_2, ..., l_n} \left( \sum_{r=0}^{d-1} \zeta^{r(l_1+l_2+...+l_n)} \right) |l_1 \oplus k_1\rangle \otimes |l_2 \oplus k_2\rangle \otimes ... \otimes |l_n \oplus k_n\rangle \\
&= d^{-\frac{n-1}{2}} \sum_{l_1+l_2+...+l_n \equiv 0 \pmod{d}} |l_1 \oplus k_1\rangle \otimes |l_2 \oplus k_2\rangle \otimes ... \otimes |l_n \oplus k_n\rangle. \quad (5)
\end{aligned}$$

If we perform quantum measurements with the $V_1$ basis on the right of Eq.(5), we will get the results of $l_i \oplus k_i$ ($i = 1,2,...,n$). According to Eq.(5), it is apparent that

$$\begin{aligned}
&(l_1 \oplus k_1) \oplus (l_2 \oplus k_2) \oplus ... \oplus (l_n \oplus k_n) \\
&= (l_1 + k_1 + l_2 + k_2 + ... + l_n + k_n) \bmod d \\
&= \left[ (l_1 + l_2 + ... + l_n) \bmod d + (k_1 + k_2 + ... + k_n) \bmod d \right] \bmod d \\
&= (k_1 + k_2 + ... + k_n) \bmod d \\
&= k_1 \oplus k_2 \oplus ... \oplus k_n. \quad (6)
\end{aligned}$$

## 2.2 Particle transmission mode of secure multi-party quantum computation

In secure multi-party quantum computation protocols (such as multi-party QKA), there are three kinds of particle transmission mode [32], i.e., the complete-graph-type, the circle-type and the tree-type (shown in Fig.1). In the complete-graph-type particle transmission mode, every party prepares the initial quantum states and sends each of the other parties a sequence of prepared particles; in the circle-type particle transmission mode, every party prepares the initial quantum states and only sends out one sequence of prepared particles which will be operated by each of the other parties in turn and finally sent back to the one who prepared it; and in the tree-type particle transmission mode, only one party prepares the initial quantum states and sends each of the other parties a sequence of prepared particles which may or may not be sent back after operation.

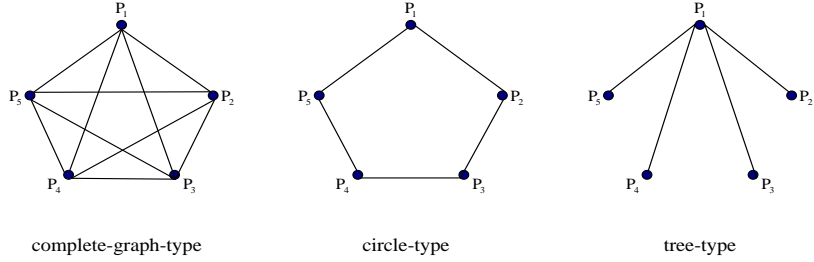

complete-graph-type    circle-type    tree-type

Fig.1 Three types of particle transmission mode in secure multi-party quantum computation protocols (taking five parties for example) [32]. Here, the vertices denote the parties while the edges denote the particle transmissions between two parties.

## 3 The proposed secure multi-party quantum summation protocol and its analysis
### 3.1 Protocol description

Secure multi-party quantum summation should meet the following requirements [52]:

① Correctness. The computation result of summation of players' inputs is correct.

② Security. An outside eavesdropper cannot obtain any useful information about each player's input without being detected.

③ Privacy. Each player cannot learn any useful information more than her prescribed out, i.e., each player's input can be kept secret.

However, the computation result of summation can be published.

Suppose that there are $n$ ($n > 2$) parties, $P_1, P_2, \ldots, P_n$, where $P_i$ ($i = 1, 2, \ldots, n$) has a private integer string $K_i$ of length $N$. That is,

$$
\begin{aligned}
K_1 &= \left(k_1^1, k_1^2, \ldots, k_1^N\right) \\
K_2 &= \left(k_2^1, k_2^2, \ldots, k_2^N\right) \\
&\vdots \\
K_n &= \left(k_n^1, k_n^2, \ldots, k_n^N\right)
\end{aligned}
\quad , \tag{7}
$$

where $k_1^t, k_2^t, \ldots, k_n^t \in \{0, 1, \ldots, d-1\}$ for $t = 1, 2, \ldots, N$. $P_1, P_2, \ldots, P_n$ want to jointly derive the summation of their private integer strings shown in Eq.(8) without revealing the genuine contents of their private integer strings.

$$
K = K_1 \oplus K_2 \oplus \ldots \oplus K_n = \left(k_1^1 \oplus k_2^1 \oplus \ldots \oplus k_n^1, k_1^2 \oplus k_2^2 \oplus \ldots \oplus k_n^2, \ldots, k_1^N \oplus k_2^N \oplus \ldots \oplus k_n^N\right). \tag{8}
$$

The detailed procedures of the proposed secure multi-party quantum summation protocol can be illustrated as follows. Without loss of generality, we suppose that $P_1$ is the party who prepares the initial quantum states. Moreover, $P_1$ is assumed to be semi-honest, which means that she may misbehave on her own but will not conspire with anyone. Here, only ideal channel (without noise) is considered.

**Step 1:** $P_1$ prepares $N$ $d$-level $n$-particle entangled states all in the state $|\omega\rangle_{12\ldots n}$, and arranges them into an ordered sequence

$$
\left[\frac{1}{\sqrt{d}}\sum_{r=0}^{d-1}|r\rangle_1^1|r\rangle_2^1\ldots|r\rangle_n^1, \ \frac{1}{\sqrt{d}}\sum_{r=0}^{d-1}|r\rangle_1^2|r\rangle_2^2\ldots|r\rangle_n^2, \ \ldots, \ \frac{1}{\sqrt{d}}\sum_{r=0}^{d-1}|r\rangle_1^N|r\rangle_2^N\ldots|r\rangle_n^N\right], \tag{9}
$$

where the superscripts $1, 2, \ldots, N$ denote the order of $d$-level $n$-particle entangled states in the sequence. Afterward, $P_1$ takes the $i^{\text{th}}$ ($i = 1, 2, \ldots, n$) particle out from each state to construct $n$ particle sequences which are labeled as:

$$S_1 = \left( p_1^1, p_1^2, \ldots, p_1^N \right)$$
$$S_2 = \left( p_2^1, p_2^2, \ldots, p_2^N \right)$$
$$\vdots$$
$$S_i = \left( p_i^1, p_i^2, \ldots, p_i^N \right), \quad (10)$$
$$\vdots$$
$$S_n = \left( p_n^1, p_n^2, \ldots, p_n^N \right)$$

where $p_i^t$ represents the $i^{th}$ particle of the $t^{th}$ entangled state and $t = 1, 2, \ldots, N$. For detecting eavesdropping, $P_1$ prepares $n-1$ groups of decoy photons, each of which is randomly chosen from the set $V_1$ or $V_2$. Then, $P_1$ randomly picks out one group of decoy photons and randomly inserts the chosen decoy photons into particle sequence $S_j$ to form a new sequence $S_j^{'}$. Here, $j = 2, 3, \ldots, n$. Finally, $P_1$ keeps $S_1$ in her hand and sends $S_j^{'}$ to $P_j$.

**Step 2:** After confirming that $P_j$ ( $j = 2, 3, \ldots, n$ ) has received all the particles in sequence $S_j^{'}$, $P_1$ checks the transmission security of sequence $S_j^{'}$ together with $P_j$. Concretely, $P_1$ tells $P_j$ the positions and the measurement basis of decoy photons in sequence $S_j^{'}$. In the following, $P_j$ uses the correct basis to measure the corresponding decoy photons and tells $P_1$ half of the measurement results. Afterward, $P_1$ announces the initial states of the remaining half of decoy photons. Finally, they check whether the measurement results of decoy photons are consistent with their initial states. In this way, $P_1$ and $P_j$ can check the transmission security of sequence $S_j^{'}$. If the error rate is greater than a predetermined threshold, they will terminate the protocol; otherwise, they will proceed to the next step.

**Step 3:** $P_j$ ( $j = 2, 3, \ldots, n$ ) discards the decoy photons in sequence $S_j^{'}$ and obtains sequence $S_j$. Then, $P_j$ encodes her private integer string $K_j$ on the particles in sequence $S_j$. Concretely, $P_j$ performs $U_{k_j^t} F$ on particle $p_j^t$, where $t = 1, 2, \ldots, N$. The new sequence of $S_j$ after encoded is denoted as $ES_j$.

In the same time, $P_1$ also encodes her private integer string $K_1$ on the particles in sequence $S_1$ by performing $U_{k_1^t} F$ on particle $p_1^t$. The new sequence of $S_1$ after encoded is denoted as $ES_1$.

**Step 4:** After all parties have finishing encoding of their private integer strings, each of them measures all particles in their respective hand with the basis $V_1$ and obtains the corresponding measurement results. As a result, it can be derived that

$$M_1 = \left( m_1^1, m_1^2, \ldots, m_1^N \right)$$
$$M_2 = \left( m_2^1, m_2^2, \ldots, m_2^N \right)$$
$$\vdots$$
$$M_i = \left( m_i^1, m_i^2, \ldots, m_i^N \right), \quad (11)$$
$$\vdots$$
$$M_n = \left( m_n^1, m_n^2, \ldots, m_n^N \right)$$

where $m_i^t$ is the measurement result of particle $p_i^t$ after encoded, $i = 1, 2, \ldots, n$ and $t = 1, 2, \ldots, N$. According to Eq.(5), it can be obtained that $m_i^t = l_i^t \oplus k_i^t$ and $l_1^t + l_2^t + \ldots + l_n^t \equiv 0 \pmod{d}$. Then, $P_j$ ( $j = 2, 3, \ldots, n$ ) announces $M_j$ to $P_1$. Finally, according to Eq.(6), $P_1$ obtains the summation of all parties' private integer strings by computing

$$M_1 \oplus M_2 \oplus \ldots \oplus M_n = \left( m_1^1 \oplus m_2^1 \oplus \ldots \oplus m_n^1, m_1^2 \oplus m_2^2 \oplus \ldots \oplus m_n^2, \ldots, m_1^N \oplus m_2^N \oplus \ldots \oplus m_n^N \right)$$
$$= \left( k_1^1 \oplus k_2^1 \oplus \ldots \oplus k_n^1, k_1^2 \oplus k_2^2 \oplus \ldots \oplus k_n^2, \ldots, k_1^N \oplus k_2^N \oplus \ldots \oplus k_n^N \right)$$
$$= K_1 \oplus K_2 \oplus \ldots \oplus K_n = K. \quad (12)$$

In order to let the other parties know the result of summation, $P_1$ announces it publicly.

It concludes the description of the proposed secure multi-party quantum summation protocol. It is apparent that in the above protocol, only $P_1$ prepares the initial quantum states and sends each of the other parties a sequence of prepared particles. Thus, the above protocol adopts the tree-type particle transmission mode.

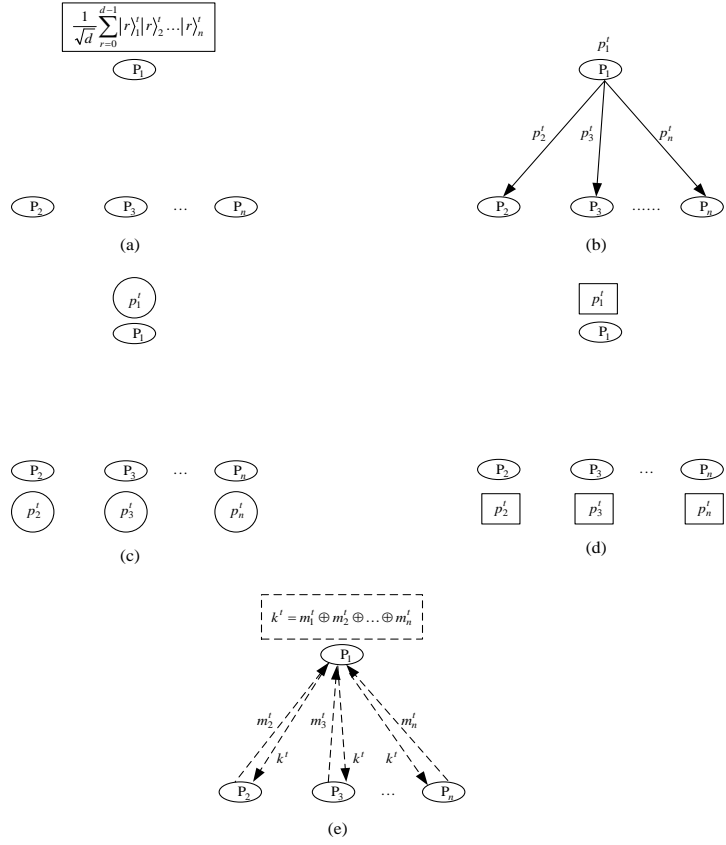

Fig.2  The flow chart of the proposed secure multi-party quantum summation protocol

(taking $\frac{1}{\sqrt{d}}\sum_{r=0}^{d-1}|r\rangle_1^t|r\rangle_2^t\ldots|r\rangle_n^t$ for example)

(a) $P_1$ prepares quantum state $\frac{1}{\sqrt{d}}\sum_{r=0}^{d-1}|r\rangle_1^t|r\rangle_2^t\ldots|r\rangle_n^t$ as the quantum carrier. Here, the rectangle with solid lines denotes the quantum state preparation operation; (b) $P_1$ transmits particle $p_j^t$ ( $j=2,3,\ldots,n$ ) to $P_j$, and keeps particle $p_1^t$ intact. Here, the solid line with an arrow denotes the quantum state transmission operation; (c) $P_i$ ( $i=1,2,\ldots,n$ ) encodes particle $p_i^t$ by performing $U_{k_i^t}F$ on it. Here, the solid circle denotes the encoding operation. (d) $P_i$ ( $i=1,2,\ldots,n$ ) measures particle $p_i^t$ after encoded with the basis $V_1$. Here, the square denotes the quantum state measurement operation. (e) $P_j$ ( $j=2,3,\ldots,n$ ) sends $m_j^t$ to $P_1$. Then, $P_1$ computes $k^t$ and sends it to $P_j$. Here, the dotted line with an arrow and the rectangle with dotted lines denote the classical information transmission operation and the classical computation operation, respectively.

## 3.2  Analysis
### A.  Output correctness

In this subsection, we verify that the output of the above protocol is correct. There are $n$ parties named $P_1, P_2, \ldots, P_n$, where $P_i$ ( $i=1,2,\ldots,n$ ) has a private integer string $K_i$ of length $N$. Without loss of generality, after ignoring the eavesdropping check processes, we take the first integer of each private integer string (i.e., $k_i^1$, $i=1,2,\ldots,n$ ) for example, to illustrate the output correctness.

$P_1$ prepares one $d$-level $n$-particle entangled state in the state $\frac{1}{\sqrt{d}}\sum_{r=0}^{d-1}|r\rangle_1^1|r\rangle_2^1\ldots|r\rangle_n^1$. Then, $P_1$ keeps particle $p_1^1$ in her hand and sends particle $p_j^1$ to $P_j$. Here, $j=2,3,\ldots,n$. After receiving particle $p_j^1$, $P_j$ performs $U_{k_j^1}F$ on particle $p_j^1$ to encode the private integer $k_j^1$. In the same time, $P_1$ also encodes her private integer $k_1^1$ by performing $U_{k_1^1}F$ on particle $p_1^1$. Then, $P_j$ measures particle $p_j^1$ after encoded with the basis $V_1$ and tells $P_1$ the measurement result $m_j^1$. $P_1$ also uses the basis $V_1$ to measure $p_1^1$ after encoded and obtains the measurement result $m_1^1$. Here, $m_i^1 = l_i^1 \oplus k_i^1$ and $i=1,2,\ldots,n$. Finally, according to Eq.(6), $P_1$ obtains $k_1^1 \oplus k_2^1 \oplus \ldots \oplus k_n^1$ by computing $m_1^1 \oplus m_2^1 \oplus \ldots \oplus m_n^1$. Concretely,

$$m_1^1 \oplus m_2^1 \oplus \ldots \oplus m_n^1 = \left(l_1^1 \oplus k_1^1\right) \oplus \left(l_2^1 \oplus k_2^1\right) \oplus \ldots \oplus \left(l_n^1 \oplus k_n^1\right)$$

$$
\begin{aligned}
&= \left(l_1^1 + k_1^1 + l_2^1 + k_2^1 + \ldots + l_n^1 + k_n^1\right) \bmod d \\
&= \left[\left(l_1^1 + l_2^1 + \ldots + l_n^1\right) \bmod d + \left(k_1^1 + k_2^1 + \ldots + k_n^1\right) \bmod d\right] \bmod d \\
&= \left(k_1^1 + k_2^1 + \ldots + k_n^1\right) \bmod d \\
&= k_1^1 \oplus k_2^1 \oplus \ldots \oplus k_n^1 = k^1 .
\end{aligned}
\tag{13}
$$

It can be concluded now that the output of the above protocol is correct.

### B. Security

In this subsection, we verify that both the outside attack and the participant attack are ineffective for the above protocol.

**(i) Outside attack**

We analyze the possibility for an outside eavesdropper to steal the private integer strings from all parties here.

In the above protocol, in order to get something useful about the private integer strings, an outside eavesdropper may utilize the particle transmission that $P_1$ sends $S_j'$ ( $j = 2,3,\ldots,n$ ) to $P_j$ in Step 1 to launch active attacks, such as the intercept-resend attack, the measure-resend attack and the entangle-measure attack and so on. However, the above protocol employs the decoy photons, which are randomly chosen from the two conjugate bases, $V_1$ and $V_2$, to detect the presence of an outside eavesdropper. Note that the decoy photon technique [62,63] can be thought as a variant of the BB84 eavesdropping check method [1] which has been proven to be unconditionally secure [64]. Moreover, the effectiveness of decoy photon technology in 2-level quantum system against an outside eavesdropper's attacks has also been validated in Refs.[65,66]. It is straightforward that the decoy photon technology is also effective against an outside eavesdropper's attacks in $d$-level quantum system. Therefore, if an outside eavesdropper launches active attacks during the particle transmissions, due to having no knowledge about the positions and the measurement basis of decoy photons before the announcement on them, she will inevitably leave her trace on decoy photons and be detected by the eavesdropping check process.

On the other hand, in Step 4, an outside eavesdropper may hear of $M_j$ when $P_j$ ( $j = 2,3,\ldots,n$ ) announces it to $P_1$ and the result of summation when $P_1$ publishes it. However, she still cannot decrypt out $k_j^t$ ( $t = 1,2,\ldots,N$ ) from $m_j^t$, because she does not know the value of $l_j^t$. In addition, an outside eavesdropper can deduce $M_1$ from $M_2, M_3, \ldots, M_n$ and the result of summation. However, due to lack of the knowledge of the value of $l_1^t$, she cannot know $k_1^t$ either.

**(ii) Participant attack**

In 2007, Gao *et al.* [67] first pointed out that the participant attack, i.e., the attack from one or more dishonest parties, is generally more powerful and should be paid more attention to. To date, the participant attack has attracted much attention in the cryptanalysis of quantum cryptography [68-70]. To see this in a sufficient way, we consider two cases of participant attack. Firstly, we discuss the participant attack from one single dishonest party; and then, we analyze the colluding attack from two or more dishonest parties.

**a) The participant attack from one single dishonest party**

In the above protocol, the roles of different $P_j$ s ( $j = 2,3,\ldots,n$ ) are the same, but are different from $P_1$ who prepares the initial quantum states and distributes the prepared particle sequences. Thus, there are two kinds of the participant attack from one single dishonest party, i.e., the participant attack from a single dishonest $P_j$ and the participant attack from semi-honest $P_1$.

With respect to the participant attack from a single dishonest $P_j$, if $P_j$ launches attacks on the particles in $S_{j'}'$ from $P_1$ to $P_{j'}$ ( $j' = 2,3,\ldots,n$ and $j' \neq j$ ) in Step 1, due to having no knowledge about the positions and the measurement basis of the inserted decoy photons in $S_{j'}'$, she will inevitably be caught as an outside eavesdropper. In addition, $P_j$ may hear of $M_{j'}$ when $P_{j'}$ announces it to $P_1$ in Step 4. However, due to having no access to the value of $l_{j'}^t$ ( $t = 1,2,\ldots,N$ ), she still cannot decrypt out $k_{j'}^t$ from $m_{j'}^t$. On the other hand, $P_j$ can deduce $M_1$ from $M_2, M_3, \ldots, M_n$ and the result of summation. However, due to lack of the knowledge of the value of $l_1^t$, $P_j$ cannot know $k_1^t$ either.

With respect to the participant attack from semi-honest $P_1$, in order to obtain the private integer strings of the other parties, $P_1$ can take the chance of preparing the initial quantum states to launch the following attack:

① $P_1$ prepares $N$ $d$-level $n$-particle entangled states all in the state $|\omega\rangle_{12\ldots n}$, and measures each of them with the basis $V_1$. The collapsed states after measurement are denoted as

$$\left[\left(\left|r^1\right\rangle_1,\left|r^1\right\rangle_2,\ldots,\left|r^1\right\rangle_n\right),\ \left(\left|r^2\right\rangle_1,\left|r^2\right\rangle_2,\ldots,\left|r^2\right\rangle_n\right),\ \ldots,\ \left(\left|r^N\right\rangle_1,\left|r^N\right\rangle_2,\ldots,\left|r^N\right\rangle_n\right)\right], \quad (14)$$

where $\left|r^t\right\rangle_i$ denotes the collapsed state of the $i^{th}$ particle in the $t^{th}$ $d$-level $n$-particle entangled state after measurement. Here, $t=1,2,\ldots,N$ and $i=1,2,\ldots,n$. Afterward, $P_1$ constructs $n$ particle sequences as follows:

$$\begin{aligned}S_1 &= \left(\left|r^1\right\rangle_1,\left|r^2\right\rangle_1,\ldots,\left|r^N\right\rangle_1\right)\\ S_2 &= \left(\left|r^1\right\rangle_2,\left|r^2\right\rangle_2,\ldots,\left|r^N\right\rangle_2\right)\\ &\vdots\\ S_i &= \left(\left|r^1\right\rangle_i,\left|r^2\right\rangle_i,\ldots,\left|r^N\right\rangle_i\right)\\ &\vdots\\ S_n &= \left(\left|r^1\right\rangle_n,\left|r^2\right\rangle_n,\ldots,\left|r^N\right\rangle_n\right)\end{aligned} \quad (15)$$

For detecting eavesdropping, $P_1$ prepares $n-1$ groups of decoy photons, each of which is randomly chosen from the set $V_1$ or $V_2$, and randomly inserts one group of decoy photons into particle sequence $S_j$ to form a new sequence $S_j^{'}$. Here, $j=2,3,\ldots,n$. Then, $P_1$ keeps $S_1$ in her hand and sends $S_j^{'}$ to $P_j$.

② $P_1$ and $P_j$ ($j=2,3,\ldots,n$) check the transmission security of sequence $S_j^{'}$ together as illustrated in Step 2. Apparently, $P_j$ cannot discover the misbehavior of $P_1$. Therefore, $P_j$ discards the decoy photons in sequence $S_j^{'}$ to restore sequence $S_j$ and performs $U_{k_j^t}F$ on particle $\left|r^t\right\rangle_j$, where $t=1,2,\ldots,N$. The corresponding encoded particle of $\left|r^t\right\rangle_j$ is

$$\left(U_{k_j^t}F\right)\left|r^t\right\rangle_j = U_{k_j^t}\frac{1}{\sqrt{d}}\sum_{l_j^t=0}^{d-1}\zeta^{l_j^t r^t}\left|l_j^t\right\rangle = \frac{1}{\sqrt{d}}\sum_{l_j^t=0}^{d-1}\zeta^{l_j^t r^t}\left|l_j^t\oplus k_j^t\right\rangle. \quad (16)$$

Afterward, $P_j$ measures all particles in her hand with the basis $V_1$ and publishes her measurement result

$$M_j = \left(m_j^1, m_j^2,\ldots,m_j^N\right). \quad (17)$$

Here, $m_j^t = l_j^t \oplus k_j^t$. Then, $P_j$ announces $M_j$ to $P_1$. Finally, $P_1$ tries to extract $k_j^t$ from $m_j^t$.

However, although $P_1$ knows $m_j^t$ from the announcement of $P_j$, she still cannot extract $k_j^t$, as she has no knowledge about $l_j^t$. It can be concluded that the participant attack from semi-honest $P_1$ is ineffective.

**b) The participant attack from two or more dishonest parties**

Since $P_1$ is not allowed to collude with other parties, if the other $n-1$ parties collude together, they can easily deduce the private integer string of $P_1$ from the result of summation. Therefore, the above protocol cannot resist the colluding attack from $n-1$ parties.

Next, we will demonstrate that the above protocol can resist the colluding attack from $n-2$ parties. Without loss of generality, assume that the dishonest $P_2,\ldots,P_{i-1},P_{i+1},\ldots,P_n$ try to collude together to obtain the private integer strings of $P_1$ and $P_i$. Firstly, if $P_2,\ldots,P_{i-1},P_{i+1},\ldots,P_n$ try to launch attacks on the particles in $S_i^{'}$ from $P_1$ to $P_i$ in Step 1, due to having no knowledge about the positions and the measurement basis of the inserted decoy photons in $S_i^{'}$, they will inevitably be caught as an outside eavesdropper. Secondly, in Step 4, $P_s$ ($s=2,\ldots,i-1,i+1,\ldots,n$) can know $M_s$, and may hear of $M_i$ when $P_i$ announces it to $P_1$ and the result of summation when $P_1$ publishes it. $P_s$ can deduce $M_1$ from $M_2,M_3,\ldots,M_n$ and the result of summation. Moreover, $P_s$ can deduce $l_s^t$ ($t=1,2,\ldots,N$) from $k_s^t$ and $m_s^t$. However, even though the $n-2$ parties conclude together, they still cannot obtain the accurate values of $l_i^t$ and $l_1^t$. Therefore, $P_2,\ldots,P_{i-1},P_{i+1},\ldots,P_n$ cannot decrypt out $k_i^t$ and $k_1^t$ from $m_i^t$ and $m_1^t$, respectively.

## 4 Discussion and conclusion

We compare the proposed protocol with previous quantum summation protocols with respect to type of addition and type of computation. The comparison result is summarized in Table 1. From Table 1, it can be concluded that the modulo of the proposed protocol can easily be bigger than those of Refs.[52,56-58,61], which may result in more extensive applications; and compared with the protocols of Refs.[52,56-58,60,61], the proposed protocol easily has higher computation efficiency because of its secret-by-secret computation.

Further, we give a more detailed comparison between the proposed protocol and the protocol of Ref.[59] by ignoring their security check processes, since both of them utilize quantum Fourier transform. The comparison result is summarized in Table 2.

Table 1 Comparison of previous quantum summation protocols and the proposed protocol

|  | The protocol of Ref.[52] | The protocol of Ref.[56] | The protocol of Ref.[57] | The protocol of Ref.[58] | The protocol of Ref.[59] | The protocol of Ref.[60] | The protocol of Ref.[61] | The proposed protocol |
|---|---|---|---|---|---|---|---|---|
| Type of addition | addition modulo 2 | addition modulo $n+1$ | addition modulo 2 | addition modulo 2 | addition modulo $N$ | addition modulo 2 | addition modulo 2 | addition modulo $d$ |
| Type of computation | bit-by-bit | bit-by-bit | bit-by-bit | bit-by-bit | secret-by-secret | bit-by-bit | bit-by-bit | secret-by-secret |

*In Ref.[56], $n$ represents the number of parties carrying a secret and $n \geq 2$; and in Ref.[59], $N=2^m$, where $m$ is the number of qubits represented by one basis state; and in the proposed protocol, $d \geq 2$.

Table 2 Comparison of quantum summation protocol in Ref.[59] and the proposed protocol

|  | Quantum resource | Quantum operation | Quantum measurement | Position for encoding a secret | Particle transmission mode | Type of addition | Type of computation |
|---|---|---|---|---|---|---|---|
| The protocol of Ref.[59] | basis state | quantum Fourier transform, controlled-not operation, oracle operation $C_j$, inverse quantum Fourier transform | $V_1$ basis measurement | global phase | circle-type | addition modulo $N$ | secret-by-secret |
| The proposed protocol | $d$-level $n$-particle entangled state | quantum Fourier transform, transformation operation $U_k$ | $V_1$ basis measurement | basis state | tree-type | addition modulo $d$ | secret-by-secret |

In addition, in some circumstance, it is necessary to make all parties share the result of summation privately among them. In other words, anyone else except all parties is not allowed to know the result of summation. In order to achieve this goal, every party can launch the proposed protocol acting as $P_1$ and does not announce the result of summation publicly.

To sum up, in this paper, a novel secure multi-party quantum summation protocol based on quantum Fourier transform is proposed, where the traveling particles are transmitted in a tree-type mode. We verify in detail that the proposed protocol can resist both the outside attacks and the participant attacks. Especially, one party cannot obtain other parties' private integer strings; and it is secure for the colluding attack performed by at most $n-2$ parties. The proposed protocol calculates the addition of modulo $d$ and implements the calculation of addition in a secret-by-secret way rather than a bit-by-bit way. In addition, the proposed protocol only considers ideal channel. When noise is concerned, additional operation such as quantum private amplification is needed.

**Acknowledgements**

The authors would like to thank the anonymous reviewers for their valuable comments that help enhancing the quality of this paper. Funding by the National Natural Science Foundation of China (Grant Nos.61402407 and 11375152) and the Natural Science Foundation of Zhejiang Province (Grant No.LY18F020007) is gratefully acknowledged.

**Compliance with ethical standards**

Conflict of interest: The authors declare that they have no conflict of interest.